\documentclass[journal=jpcbfk,manuscript=article]{achemso}

\usepackage{graphics}
\usepackage[dvips]{color}
\usepackage[T1]{fontenc}
\usepackage{graphicx}
\usepackage{amssymb}
\usepackage{rotating}
\usepackage{epsfig}

\author{Tamio Yamazaki}
\affiliation{Simulation $\&$ Analysis R$\&$D Center, Canon Inc. 3-30-2, Shimomaruko,
 Ohta-ku, Tokyo 146-8501, Japan.}
\email{yamazaki.tamio@canon.co.jp}

\title
  {Breakdown of time-temperature superposition in a bead-spring polymer
 melt near the glass transition temperature}

\begin{document}

%%%%%%%%%%%%%%%%%%%%%%%%%%%%%%%%%%%%%%%%%%%%%%%%%%%%%%%%%%%%%%%%%%%%%
%% The "tocentry" environment can be used to create an entry for the
%% graphical table of contents. It is given here as some journals
%% require that it is printed as part of the abstract page. It will
%% be automatically moved as appropriate.
%%%%%%%%%%%%%%%%%%%%%%%%%%%%%%%%%%%%%%%%%%%%%%%%%%%%%%%%%%%%%%%%%%%%%
%%\begin{tocentry}
%%
%%Some journals require a graphical entry for the Table of Contents.
%%This should be laid out ``print ready'' so that the sizing of the
%%text is correct.
%%
%%Inside the \texttt{tocentry} environment, the font used is Helvetica
%%8\,pt, as required by \emph{Journal of the American Chemical
%%Society}.
%%
%%The surrounding frame is 9\,cm by 3.5\,cm, which is the maximum
%%permitted for  \emph{Journal of the American Chemical Society}
%%graphical table of content entries. The box will not resize if the
%%content is too big: instead it will overflow the edge of the box.
%%
%%This box and the associated title will always be printed on a
%%separate page at the end of the document.
%%
%%\end{tocentry}
%%
%%%%%%%%%%%%%%%%%%%%%%%%%%%%%%%%%%%%%%%%%%%%%%%%%%%%%%%%%%%%%%%%%%%%%
%% The abstract environment will automatically gobble the contents
%% if an abstract is not used by the target journal.
%%%%%%%%%%%%%%%%%%%%%%%%%%%%%%%%%%%%%%%%%%%%%%%%%%%%%%%%%%%%%%%%%%%%%
\begin{abstract}
The breakdown of the time-temperature superposition (TTS) near its glass transition temperature ($T_g$) in simple bead-spring polymer melts with and without the chain angle potential was numerically investigated. The stress relaxation modulus at different temperatures $G(t,T)$ was calculated by the Green-Kubo relation. The TTS of $G(t,T)$ of bead-spring polymer melts worked well at temperatures sufficiently higher than its $T_g$. However, when the system temperature is approaching the glass transition regime, the breakdown of TTS is observed. At temperatures near the $T_g$, the temperature dependence of the shift factor (${\rm{a}}_T^B$), which is defined at the time scale between the bond relaxation and the chain relaxation regimes of a $G(t)$-function, is significantly stronger than ones (${\rm{a}}_T^A$) defined by the time scale of the chain relaxation modes. In direct relation to the breakdown of TTS of $G(t,T)$, the decoupling of Stokes-Einstein law of diffusion-viscosity relation also appears in the glass transition regime. The analysis of the van Hove function $G_{s}(r,t)$ and non-gaussian parameter ,$\alpha_{2}\left(t\right)$, of the bead motions strongly suggest that the TTS breakdown is concerned with the dynamic heterogeneity. The effect of the chain stiffness on the temperature dependence of the shift factors was also investigated in this study. The stiffer chains melt has a stronger temperature dependence of the shift factors than the ones of the flexible chains melt. However, regardless of the chain stiffness, the stress relaxation modulus functions of the bead-spring polymer melts will begin to breakdown the TTS at a similar $T_g$-normalized temperature around $T/T_g \approx 1.2 $.
\end{abstract}

%%%%%%%%%%%%%%%%%%%%%%%%%%%%%%%%%%%%%%%%%%%%%%%%%%%%%%%%%%%%%%%%%%%%%
%% Start the main part of the manuscript here.
%%%%%%%%%%%%%%%%%%%%%%%%%%%%%%%%%%%%%%%%%%%%%%%%%%%%%%%%%%%%%%%%%%%%%
\section{Introduction}
\hspace*{1em}Amorphous polymer materials are commonly used in various industrial products, such as packaging films, body materials, electro-photographic toners, adhesive agents, buffer materials, etc. From the viewpoints of the control of the viscoelastic and mechanical properties of these amorphous polymer materials, understanding of the dynamics of the polymer chain is one of the essentials of polymer science and engineering.
The time-temperature superposition (TTS) principle is a useful concept widely used for the analysis of the dynamic properties of a polymer.\cite{Daly2009}
 According to the time-temperature superposition, which assumes that all of the relaxation modes of a polymer chain obey the same temperature dependence, the dynamic material functions obtained at several different temperatures can be collapsed into a "master" curve by shifting the time scale of each function.
The shifting coefficients, which are the so-called "shift factor", are represented as a function of the temperature of the system. The dependence of the shift factor on temperature can be well described by the Vogel-Fulcher-Tammann (VFT) or the Williams-Landel-Ferry (WLF) equation, both being basically the same equation, and the equations are applicable above the glass transition temperature ($T_g$) + ca. 50 K.
However, in the case when the system temperature approaches the $T_g$, the breakdown of TTS can be observed in many polymer systems.\cite{Ferry1980,Plazek1965}
Incidentally, the breakdown of TTS of the dynamic material functions lead to the decoupling of Stokes-Einstein law of diffusion-viscosity relation at low temperature.\cite{Stillinger2001}
Using the viscoelastic and optical birefringence measurements of polystyrene, Inoue and co-workers \cite{Inoue1999} revealed that the stress relaxation of a polymer melt has two components. One (R-component) is related to the relaxation of the orientation of a polymer chain, which is well described by the Rouse theory, and the other (G-component) is related to the relaxation of the transverse component of the monomeric motion on a length scale shorter than the shortest Rouse mode. These two components have different temperature dependencies to each other.
 The G-component has the strongest temperature dependence versus that of the R-component.
Obviously, the existence of two relaxation modes having different temperature dependencies can be due to the breakdown of the TTS of the viscoelastic functions. Similarly, the TTS of the dielectric relaxation also breaks down at a temperature close to the glass transition temperature for the polymer. The broadband dielectric relaxation measurements of amorphous polymer melts\cite{Roland1996,Plazek2001,Sokolov2006,Sokolov2007} show a significant difference between the temperature dependencies of the relaxation times of the segment mode and the normal mode. The normal mode is a relaxation behavior due to fluctuation of the end-to-end vector of a polymer chain and the segment mode is one of several beaded monomer units. The ratio of the relaxation times between the normal mode ($\tau_n$) and segmental mode ($\tau_s$) is constant, as long as the TTS can be applicable at a sufficiently higher system temperature than its $T_g$, though the ratio ($\tau_n/\tau_s$) gradually decreases when the system temperature is approaching the $T_g$. The relaxation time of the segmental mode has a stronger temperature dependence than the ones of the normal mode.
Several lines of studies suggested that the difference in the temperature dependence between the chain and segmental relaxation times is due to dynamic heterogeneities at a temperature close to the $T_g$ of the polymer.\cite{Sokolov2009,Harrowell2012} However, the basic mechanism for the thermo-rheological complexity is still not completely understood.\cite{Roland2013} 
This investigation has been highly motivated by these studies concerned with the TTS of polymer dynamics near the $T_g$.
 In this study, the applicability of TTS to the stress relaxation modulus functions $G (t,T)$ of dense bead-spring polymer melts over a wide temperature and time range were evaluated. The effect of the chain rigidity in a polymer melt on its relaxation behavior in the glass-rubber transition regime will also be discussed in this study. 

\section{Method}

\subsection{A) Simulation model}
  In this section, the details of the simulation model and computational methods for the dense bead-spring polymer melt are described. The polymer molecules are represented by the soft-core spheres and the stretch and bending springs. Two types of polymer chains were prepared; one is the freely-jointed chain ($FJC$), which has a stretching spring between neighboring beads, while the other is the freely-rotating chain ($FRC$), which is the extension model of $FJC$ to describe the stiffness of the chain. $FRC$ has both stretching and bending springs. All monomers interact through the Lennard-Jones (12-6) potential  
\begin{equation}
 U_{\rm{non}}(R_{ij})=4 \epsilon \left(\left(\frac{\sigma}{R_{ij}}\right)^{12}
        -\left(\frac{\sigma}{R_{ij}}\right)^{6}\right),
\end{equation}
where $R_{ij}$ is the distance between two beads, $\sigma$ is the (finite) distance at which the inter-particle potential is zero, and $\epsilon$ is the depth of the potential well. $U_{\rm{non}}(R_{ij})$ is truncated at $R_{ij}$ = 2.0 $\sigma$. In addition, the bonded neighbors in a chain interact through the FENE bond potential
\begin{equation}
 U_{\rm{bond}}(R_{ij})=-15R_{0}^{2}{\rm{ln}}[1-(R_{ij}/R_{0})^2] ,
\end{equation}
where $R_{0}$ is the maximum length of the bond. The equilibrium bond length $l_{0}$ =0.96$\sigma$ with $R_{0}$ = 1.5$\sigma$. For the $FRC$, the harmonic angle potential 
\begin{equation}
 U_{\rm{angle}}(\theta)= k_{\theta}[1-\rm{cos}\theta] ,
\end{equation}
where $\theta$ is the external angle between two adjacent bonds and $k_{\theta}$ is the force constant ($k_{\theta}$ = 2.0${\epsilon}$ ), is added to the total energy. The simulation box with the periodic boundary condition includes 133 polymer chains of length N = 30, and has 3990 particles in total. All the $NpT$ simulations of this study are performed by the cognac 8.0 codes.\cite{Aoyagi2002} The production time step for integration is dt = 0.01$\tau$, where $\tau$ is the unit of time $\tau=(m{\sigma}^{2}/{\epsilon})^{0.5}$, and $m$ is the mass of the bead ($m=1$ in this study). The Nose-Hoover thermostat \cite{Nose_Hoover1,Nose_Hoover2,Nose_Hoover3} and Andersen barostat \cite{Andersen1980} are used to control the temperature $T$ and the pressure $P$ of the system, respectively. For all simulations in this study, the pressure of the system was set to zero ($P=0$)

\subsection{B) Measurement of the glass transition temperature($T_{g}$) }
In this study, $T_{g}$ is measured from the temperature dependence of the specific volume of a polymer melt. The specific volume of a polymer melt decreases with the decreasing temperature, but the shrinkage rate of the volume of the polymer melt changes to a lower value at a temperature close to $T_{g}$. It was noted that such a volume change in the system will continuously occur, in contrast to the abrupt change in volume at the freezing point of a crystalline material. $T_{g}$ can be defined as the intersection of two regression lines of the specific volume versus the temperature in the liquid state and in the glassy state. The molecular dynamics simulations of the dense bead-spring melts with the step-wise cooling from $T = 1$ to $T = 0.14$ with a rate of ${\Delta} T = 0.02$ per ${\Delta} t = 2000{\tau}$ was performed. The cooling rate $\Gamma$ is $10^{-5}{\tau}^{-1}$. 
 Universally, $T_g$ depends on $\Gamma$, the $T_g$ at low $\Gamma$ is lower than ones at high $\Gamma$, $\Gamma$-dependence on $T_g$ for a dense bead-spring melt is reported by Buchholz et.al.\cite{Buchholz2002}, which exhibit that $T_g(\Gamma)$ is almost constant at $\Gamma < 10^{-3} {\tau}^{-1}$, although $T_g$ is significantly increasing with the increasing of $\Gamma > 10^{-1} {\tau}^{-1}$. 
It is expected that the cooling rate of $\Gamma = 10^{-5} \tau$ is the sufficiently small value for the estimation of the asymptotic value of $T_g(\Gamma)$ for $FJC$ and $FRC$ melts. 

\subsection{C) Calculation of the relaxation modulus $G(t)$}
$G(t)$ is obtained using Green-Kubo formula,
\begin{equation}
 G(t)=\frac{V}{{\epsilon}T}{\langle{{\sigma_{xy}}\left({t}\right)}
                                   {{\sigma_{xy}}\left({0}\right)}\rangle} ,
\end{equation}
 where $V$ is the volume of the system,$\sigma_{xy}(t)$ is the stress tensor of the system given by 
\begin{equation}
\sigma_{xy}\left({t}\right) = -\frac{1}{V}\left(\sum_{i}^{N}{m_{i}v_{i,x}v_{i,y}} + \sum{i}\sum_{j>i}^{N}{R_{ij,x}f_{ij,y}}\right),
\end{equation}

where $N$ is the total number of beads in the system, $m_{i}$ is the mass of a bead, $v_{i,x}$ and $v_{i,y}$ are the $x$-component and $y$-component of the velocity vector of the $i$-th bead respectively, $R_{ij,x}$ is the $x$-component of position vector from $i$-th bead to $j$-th bead, and $f_{ij,y}$ is the y-component of force acting between the $i$-th and $j$-th beads. In order to obtain an accurate $G(t)$, the stress tensor should be calculated at every time step. However, it is inefficient to store and access such huge files of the values of the instantaneous stress components. To avoid this difficulty, the correlator algorithm described in Ref.~\citenum{Likhtman2007} was used. The algorithm utilizes the instantaneous stress values at every MD step for the calculation of $G(t)$ function, although using an array of correlators and an accumulator in this algorithm require only a small memory size to store the data. 

\section{Results and discussion} 
\subsection{A) Glass transition temperature} 
The temperature dependences of the specific volume of two bead-spring polymer melts ($FJC$ and $FRC$) are shown in Figure 1. The plotted volume data are the mean values from five independent MD simulations. It is noted that there is no difference in the specific volume between the $FJC$ and $FRC$ melts at temperatures above the glass transition region. However, the glass transition region of $FRC$ melt begins at a higher temperature than that of the $FJC$ melt. Both curves have an obvious bending point. The glass transition temperature was defined as the intersection point of two regression lines obtained from the lower and higher temperature regions than the bending point, and these obtained values are $T_{g}$ = 0.44 (for $FJC$) and $T_{g}$ = 0.56 (for $FRC$). 

\begin{figure}[h] 
\centering
  \includegraphics[height=10cm]{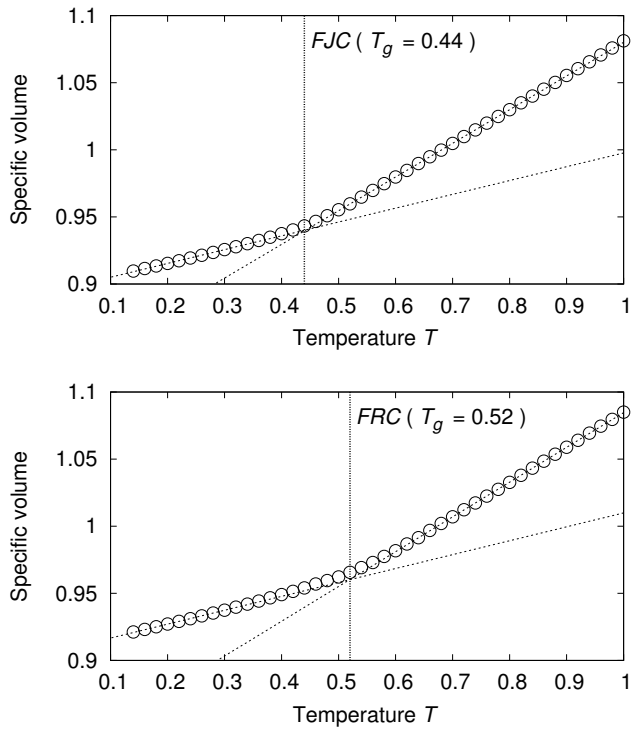}
  \caption{Specific volume ($V_{sp}$) of $FJC$ melt(upper) and $FRC$ melt (bottom) as a function of temperature at the cooling rate of $10^{-5}\tau^{-1}$. The open circles represent the simulated volume data, and the dashed lines represent the regression results individually fitted to the volume data at temperatures lower and higher than the bending point of the function. The plotted data are obtained from five independent MD runs of 2000 $\tau$.}
  \label{fgr:example}
\end{figure}
\clearpage

\subsection{B) Determination of the shift factors for $G(t)$}
As mentioned in the Introduction, the dynamic materials functions (e.g., a relaxation modulus, a diffusion coefficient, viscosity, creep compliance, etc.) at several different temperatures can be coincided with a master curve by shifting the time scale of each function. The TTS can be applied to the $G(t)$ in the following manner, 
\begin{equation}
{\rm{b}}_{T}^{-1}(T) G\left({T,t/{\rm{a}}_{T}(T)}\right) = G(T_{0},t),  
\end{equation}
where $T_0$ is the reference temperature ($T_0$ = 1 in this study) and ${\rm{b}}_{T}(T)$ and ${\rm{a}}_{T}(T)$ are the vertical and horizontal shift factors, respectively. The ${\rm{b}}_{T}(T)$ value is given by 
\begin{equation}
{\rm{b}}_{T}(T)=\frac{\rho\left({T}\right){T}}{\rho\left({T_0}\right){T_0}}, 
\end{equation}
\\
where $\rho\left({T}\right)$ is the density of the polymer melt at the temperature $T$. Eq.7 is derived based on the analogy of the classical rubber elasticity theory. In order to discuss the time-dependence of the horizontal shift factor, two shift factors(${\rm{a}}_{T}^{A}$ and ${\rm{a}}_{T}^{B}$) were defined at two different times ($\tau_A$ and $\tau_B$) as follows:
\begin{eqnarray} 
G\left({T,t/{\rm{a}}_{T}^{A}(T)}\right)={\rm{b}}_{T}(T)G(T_{0},\tau_A),  \nonumber \\
G\left({T,t/{\rm{a}}_{T}^{B}(T)}\right)={\rm{b}}_{T}(T)G(T_{0},\tau_B),  \nonumber \\
\end{eqnarray}
 where $\tau_A$ and $\tau_B$ are chosen at around the onset ($\tau_B$ = 1 $\tau$) and middle ($\tau_A$ = 100 $\tau$) of the time-range where the chain-relaxation is observed in the $G(t)$ function at $T_0$ = 1. 
$\tau_A$ is roughly equivalent to the time-scale of the relaxation of three or four segments, and $\tau_B$ is shorter than the relaxation time-scale of a dimer. Due to the large fluctuations in the $G(t)$ functions, the shift factors $({\rm{a}}_T)$ are chosen to maximize the number of plotted points of ${\rm{b}}_T^{-1}G(T,t/{\rm{a}}_T)$ in the windows which are located at the coordinates of the points of $G(T_0,t)$ around $t\approx{\tau_A}$ or ${\tau_B}$. Although five squares with the side length of 0.1 (on a double logarithmic scale) were employed, the choice of these parameters does not significantly affect the TTS results.
Figure 2 shows the stress relaxation modulus $G(t,T=1)$ of the $FJC$(black line) and $FRC$(gray line) polymer melts in the log-log plot of $G(t)$ versus time, which are averages of ten independent MD-runs. The oscillations at an early time ($t$ < 1 $\tau$ ) are due to the bond relaxation of a chain. The slope of $G(t)$ versus time on the log-log scale at $t$ = $\tau_A$ is about -0.7 (for $FJC$) and -0.5 (for $FRC$). The Rouse prediction value is -1/2. The curve of $G(t)$ for the $FRC$ melt is somewhat extended in comparison to the one for $FJC$. The stretching of $G(t)$ in the $FRC$ melt suggests a polymer chain entanglement, which is not observed in the $FJC$ ($N$ = 30) melt, that can occur due to the rigidity of a polymer chain. \cite{KG1990}
The vertical shifted stress relaxation moduli (${\rm{b}}_{T}(T)^{-1}G(t,T)$) of the $FJC$ and $FRC$ melts at several different temperatures are shown in Figure 3 ($FJC$) and Figure 4 ($FRC$), respectively. Each curve of ${\rm{b}}_{T}(T)^{-1}G(t,T)$ is an average of ten independent MD runs. The MD runs with the length of $10^{6}{\tau}$ are performed at each temperature (varied from $T$=1 to near its $T_{g}$). The pressures of all the systems treated in this study were set to zero. 
${\rm{b}}_{T}(T)^{-1}G(t,T)$ of the chain-relaxation regime is shifted to the right hand with the decreasing temperature, on the other hand, one of the bond-stretching relaxation regime migrates to up. Therefore, the horizontal width between the end of a bond stretching relaxation regime and the onset of a chain-relaxation regime is gradually extended as the temperature decreases. The relaxation modes in this crossover regime are not clearly assigned in this study, though Likhtman et al. expressed them as the "colloidal or glassy modes" in Ref.\citenum{Likhtman2007}. The middle graph(b) in Figs. 3 and 4 shows the results of the TTS reduced to $T_{0}$ = 1 by the definition (Eq.8) of ${\rm{a}}_T^A$. The TTS works well at a much higher temperature compared to its $T_g$. However, at lower temperatures (near $T_g$), the $G(t,T)$ functions at different temperatures will not collapse into one single universal curve. Obviously, a different relaxation process, which has a different temperature dependence, appeared at around ${\tau_B}$ ($t/{\rm{a}}_{T}(T) = 1 {\tau}$). Instead of using the shift factor ${\rm{a}}_T^A$, the results of the TTS using the shift factors ${\rm{a}}_{T}^{B}$ defined by Eq.8, are shown (bottom graph) in Figure 3 ($FJC$) and in Figure 4 ($FRC$). Similar to the middle graph (shifted by ${\rm{a}}_T^A$), the curves can be collapsed to $G(T_0,t)$ at a higher temperature (above $T_g$), conversely, they deviate from $G(T_0,t)$ at a lower temperature (near $T_g$).  

\begin{figure}[h] 
\centering
  \includegraphics[height=5cm]{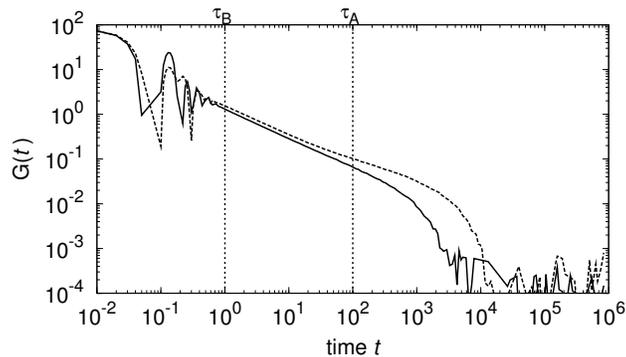}
  \caption{The stress-relaxation modulus at $T = 1$  and $P = 0$ : the solid line ($FJC$ melt) and dashed line ($FRC$ melt). The vertical lines indicate the times ($\tau_A$ and $\tau_B$) at which the horizontal shift factors (${\rm{a}}_T^A$ and ${\rm{a}}_T^B$ ) are defined.}
  \label{fgr:example}
\end{figure}
\clearpage
\begin{figure}[h] 
\centering
  \includegraphics[height=15cm]{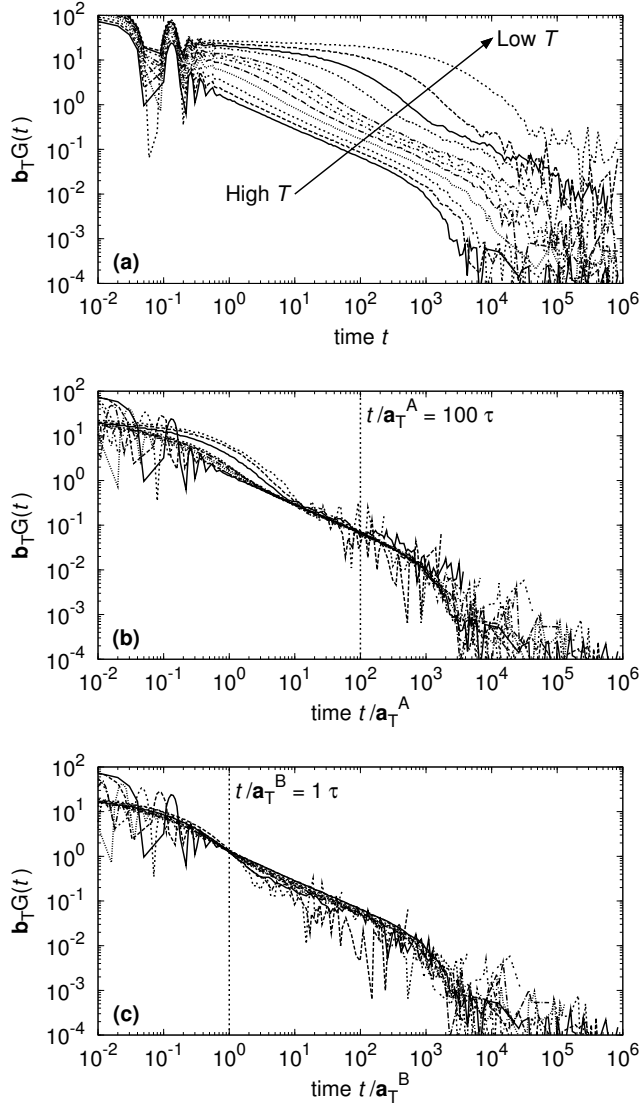}
  \caption{The stress-relaxation modulus with a vertical shift factor (${\rm{b}}_T^{-1}G$) of $FJC$ melt as a function of time at different temperatures ($T=1, 0.9, 0.8, 0.7, 0.64, 0.6, 0.58, 0.56, 0.52, 0.5, 0.48,$ and$ 0.46$). The upper graph (a) shows the ${\rm{b}}_T^{-1}G(t)$ without any horizontal shift factor. The middle (b) and bottom (c) graphs show the time-temperature superposition of ${\rm{b}}_T^{-1}G(t)$ using the horizontal shift factors ${\rm{a}}_T^A(T)$ and ${\rm{a}}_T^B(T)$, respectively.  The plotted data are obtained from ten independent MD-runs. The vertical dashed lines in (b) and (c) indicate the times, at which the horizontal shift factors ( ${\rm{a}}_T^A(T)$ and ${\rm{a}}_T^B(T)$ ) are defined.}
  \label{fgr:example}
\end{figure}
\clearpage
\begin{figure}[h] 
\centering
  \includegraphics[height=15cm]{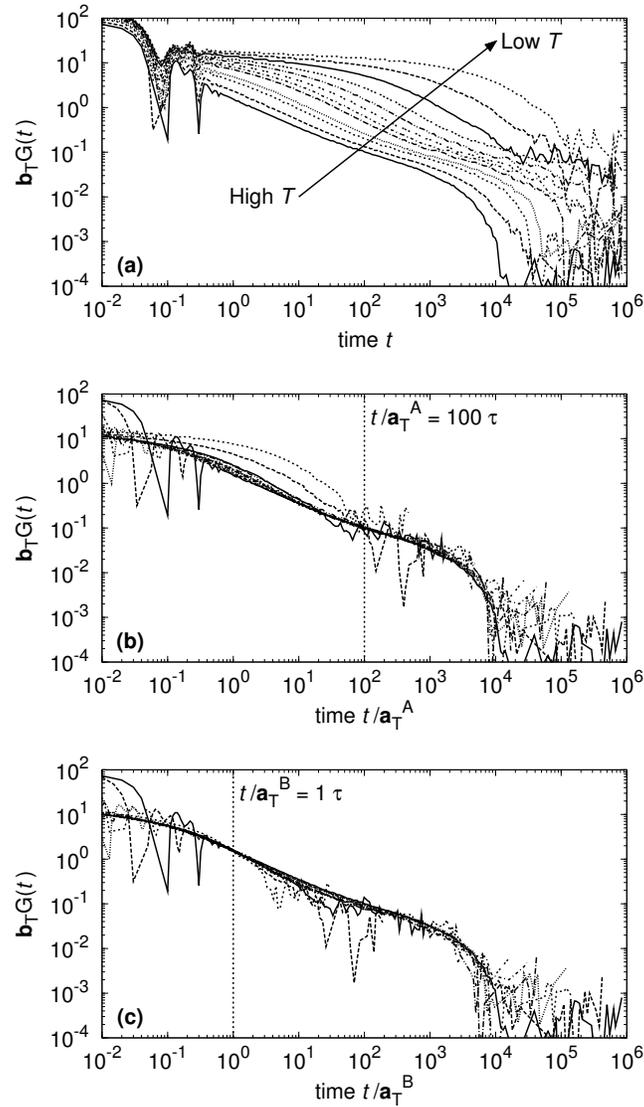}
  \caption{The stress-relaxation modulus with a vertical shift factor (${\rm{b}}_T^{-1}G$) of $FRC$ melt as a function of time at different temperatures ($T=1, 0.9, 0.8, 0.76, 0.70, 0.68, 0.66, 0.64, 0.62, 0.6, 0.58,$ and $0.56$ ).  The same format as shown in Figure 3 is used. 
}
\label{fgr:example}
\end{figure}
\clearpage

\subsection{C) Temperature dependence of the shift factors }
Figure 5 shows the horizontal shift factors, "${\rm{a}}_T^A(T)$" and "${\rm{a}}_T^B(T)$", of the $FJC$(upper) and $FRC$(bottom) melts as a function of the $T_g$-normalized temperature. Solid lines in Figure 5 are the least squares fits to the data of shift factors using WLF equation given by: 
\begin{equation}
{\log}{\rm{a}_{T}}\left(T\right) = -\frac{c_1\left({T-T_0}\right)}{2.303\left({c_2+\left({T-T_0}\right)}\right)},
\end{equation}
where $c_1$ and $c_2$ are WLF parameters, and $T_0$ is the refarence temperature ($T_0 = 1$). 
The WLF parameters were obtained for $FJC$ as $c_1=1.57, c_2=0.65$ ($\rm{a}_{T}^{A}$), $c_1=1.75, c_2=0.64$ ($\rm{a}_{T}^{B}$), and for $FRC$ as $c_1=2.69, c_2=0.59$ ($\rm{a}_{T}^{A}$), $c_1=2.35, c_2=0.54$ ($\rm{a}_{T}^{B}$). 
At sufficiently higher temperatures, above its $T_g$, ${\rm{a}}_T^A(T)$ is in total agreement with ${\rm{a}}_T^B(T)$ for both polymer types ($FJC$ and $FRC$). At low temperatures near its $T_g$, ${\rm{a}}_T^B(T)$ will become greater than ${\rm{a}}_T^A(T)$ and the rate of ${\rm{a}}_T^A(T)$ by ${\rm{a}}_T^B(T)$ decreases with the decreasing of the temperature, as shown in Figure 6. The extrapolation to $T/T_g$ = 1 for the line in Figure 6 estimates that the ${\rm{a}}_T^A(T)/{\rm{a}}_T^B(T)$ of $FJC$ could vary by an order of magnitude or even more. The ratios of ${\rm{a}}_T^A(T)/{\rm{a}}_T^B(T)$ of the $FRC$ melt are also shown in Figure 6 by the open triangles.
The temperature dependence of ${\rm{a}}_T^A(T)/{\rm{a}}_T^B(T)$ of the $FRC$ is greater than the ones for $FJC$, and vary by a greater order of magnitude (or even more) than the ones for $FJC$ by extrapolating to $T/T_g$ = 1. However, both temperatures normalized by the $T_g$ at the beginning of the decrease of ${\rm{a}}_T^{A}(T)/{\rm{a}}_T^{B}(T)$ for the $FJC$ and $FRC$ melts are observed around $T/T_g$ = 1.2. This may mean that the $T_g$-normalized temperature at which ${\rm{a}}_T^A(T)/{\rm{a}}_T^B(T)$ begin to decrease (i.e., begins the breakdown of TTS ) is independent of the rigidity of the chain. This indifference to the stiffness of the chain can also be seen in the temperature dependencies of the ratio of the segmental (${\tau_s}$) and chain(${\tau_n}$) relaxation times which are observed by the dielectric relaxation measurements.
The rate ${\tau_n/\tau_s}$ of several well-defined homo-polymers (PIP,PP,PPG,PC and POG) starts to drop in a similar range of ${\tau_s}$ ($\tau_s =10^{-5}$ to $10^{-7}{\rm{s}}$), although a more rigid polymer chain will lead to a greater temperature dependence of ${\tau_n}/{\tau_s}$. \cite{Sokolov2006}
 $T_g$ is often defined as $\tau_s(T_g)=100s$ \cite{Roland1998}, thus there is the huge amount of change ($\tau_s = 10^{2}$ to $10^{-12}s$) of $\tau_s$ in the glass transition regime. The temperature $T$, at which $\tau_s(T) = 10^{-5}$ to $10^{-7} {\rm{s}}$ corresponds to $T \approx 1.2 T_g$ for fragile glass formers\cite{Roland2003,Schwartz2006,Kunal2008}, which are applicable to many amorphous polymers. 
 It is surprising that such simple bead-spring model (linear topology, without a side-chain group) investigated in this study exhibits the temperature-dependencies of the shift factors on short and long time scales, which are qualitatively comparable with the experimental observations. 

\begin{figure}[h] 
\centering
  \includegraphics[height=10cm]{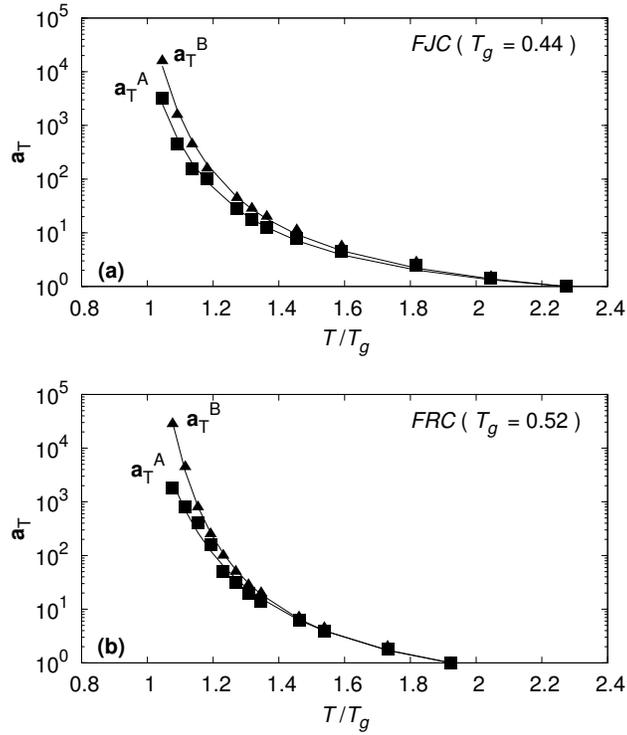}
  \caption{The $T_g$-scaled temperature dependence of horizontal shift factors(${\rm{a}}_{T}^{A}$ and ${\rm{a}}_{T}^{B}$) of $FJC$ melt (upper) and $FRC$ melt (bottom). The filled squares represent the logarithmic value of the ${\rm{a}}_{T}^{B}$, and the gray filled circles represent the ones of ${\rm{a}}_{T}^{A}$. Solid lines are least squares curves fit to the WLF equation (Eq.9).}
  \label{fgr:example}
\end{figure}
\begin{figure}[h] 
\centering
  \includegraphics[height=6cm]{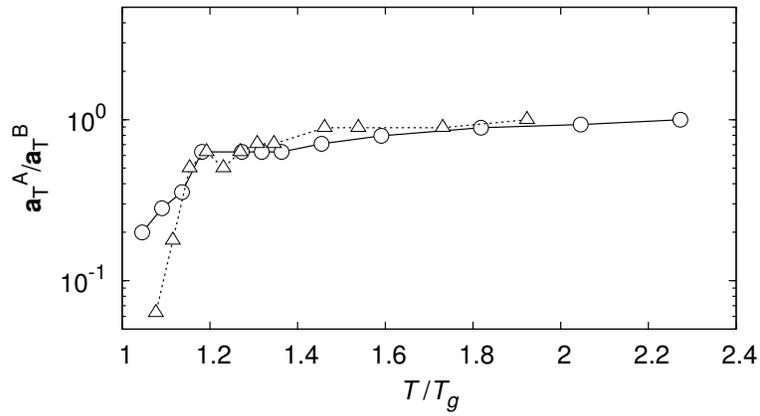}
  \caption{The $T_g$-scaled temperature dependence of the ratio (${\rm{a}}_T^{A}(T)/{\rm{a}}_T^{B}(T)$): the open circles represent the values of the $FJC$ melt, and the open triangles represent the values of the $FRC$ melt.Solid and dot lines are drawn point-to-point with a straight line with no smoothing.}
  \label{fgr:example}
\end{figure}
\clearpage

\subsection{D) Temperature dependence of diffusion and viscosity}
The self diffusion coefficient ($D$) of a molecule in a liquid is affected by the viscosity($\eta$) of the liquid. According to Stokes-Einstein law, $D{\eta}T^{-1}$ should be constant with temperature. However, a decoupling between the diffusion and the viscosity due to the dynamic heterogeneity near $T_g$ is reported in the experimental\cite{Swallen2003,Urakawa2004,Roland2004} and theoretical\cite{Yamamoto1998,Varnik2002} studies on many supercooled liquid systems. To examine the relationship between the decoupling of $D$-$\eta$ relation and the breakdown of TTS of $G(t,T)$, the temperature dependencies of the self diffusion coefficient of a chain and of the viscosity of $FJC$ melt are additionally evaluated in this study. The self diffusion coefficient of a chain is calculated from the mean square displacement  of the center of mass ($g_3(t)$) of a chain as follows:
\begin{equation}
{g_{3}}(t)=\left<{\frac{1}{M}}{\sum_{j=1}^{M}{ \left| {{\rm{r}}_{j}^{\rm{com}}(t)-{\rm{r}}_{j}^{\rm{com}}(0) }\right|^{2}} }\right>, 
\end{equation}
where ${\rm{r}}_{j}^{\rm{com}}$ is position vector of the center of mass of $j$-th chain, $M$ is the total number of chains in the system. 
By using of Einstein relation, $D$ is obtained as:
\begin{equation}
D=\lim_{t \to \infty}{\frac{g_3(t)}{6t}}.
\end{equation}
$D$ at different temperatures are determined using each data of $g_3(t)$ during $t=1{\times}10^{5}{\tau}$ to $5{\times}10^{5}$. 
On the other hand, the zero-shear viscosity $\eta$ of a liquid is given by:
\begin{equation}
{\eta}(T) = \int_{t=0}^{t=\infty}{G(t,T)}dt = \int_{t=0}^{t=t_d}{G(t,T)}dt + \int_{t=t_d}^{t=\infty}{G(t,T)}dt,
\end{equation} 
where $t_d$ is a positive number, and $G(t,T)$ denotes a stress relaxation modulus. Due to the large fluctuation or the shortage of data of the long time scale part of $G(t)$ obtained by MD simulations, it is difficult to evaluate ${\eta}$ with Eq.11. And so, the long part of $G(t)$ is approximated by the stress relaxation modulus function based on Rouse model, and ${\eta}(T)$ can be rewritten as: 
\begin{equation}
{\eta}(T) \approx \int_{t=0}^{t=t_d}{G(t,T)}dt + {\frac{{\rho}T}{N}}\int_{t=t_d}^{t=\infty}{\sum_{p=1}^{N-1}}{\left({\frac{1}{p^2}}\right)}{\rm{exp}}\left({-\frac{2t}{p^2\tau_R}}\right)dt,
\end{equation}  
where $p$ is an integral number from 1 to $N-1$, $\tau_R$ is Rouse relaxation time of a chain. Assuming the parfect TTS of $G(t)$ at long time scale $t/{\rm{a_T^A}}>\tau_A$ (with $\tau_A$ = 100 $\tau$), $t_d$ is chosen as $t_d/{\rm{a_T^A}}=\tau_A$, and $\tau_R$ has the same temperature dependence at ${\rm{a_T^A}}$ and is expressed as $\tau_R(T) = {\rm{a_T^A}} \tau_R(T_0)$. $\tau_R(T_0)$ is the Rouse relaxation time at the refarence temperature ($T_0 = 1$),which is estimated as $\tau_R(T_0) \sim 1500 \tau $, by fitting of Rouse model to the data of the long time part of $G(t,T_0)$. Cnsequently, the value of the second term of the right-hand of Eq.13 is obtained as $26.6{\cdot}{\rm{a_T^A}}{\cdot}{\rho}{\cdot}{T}$. The first term of right-hand of Eq.13 is calculated by numerical integration of the $G(t)$ vs. $t$ data. Figure 7 shows that temperature dependencies of $1/D$ and $\eta/T$ of FJC melt. The $1/D$ and $\eta/T$ are divied by the corresponding value of each quanity at reference temperature $(T_0 = 1)$. 
These data of $D(T_0)/D(T)$ and $\eta(T)T^{-1}/\eta(T_0)T_0^{-1}$ can also be fit to the WLF equation (Eq.9), and the WLF parameters were determined for $FJC$ melt as $c_1$ = 1.70, $c_2$ = 0.66 ($\log[D(T_0)/D(T)]$) and $c_1$ = 1.62, $c_2$ = 0.65 ($\log[\eta(T)T^{-1}/\eta(T_0)T^{-1}]$), respectively. 
Though, $1/D$ and $\eta/T$ exhibit the same themperature dependence at temperatures sufficiently higher than its $T_g$, $\eta/T$ has the somewhat greater themperature dependence than ones of $1/D$ when the system temperature approaches its $T_g$ ($T<1.2T_g$).This means that the breakdown of Stokes-Einstein law of viscosity-diffusion relation for $FJC$ melt occurs in the glass transition regime. These trands of temperature dependence of $D$ and $\eta/T$ calculated in this study, are consistent with the experimentally observed ones of low molecular weight unentangled polystyrene\cite{Urakawa2004,Roland2004}. Although, the effect of the swelling of $G(t)$ on the viscosity is greatly attenuated by the contribution of Rouse modes at long time scale, the difference between $\eta(T)T^{-1}/\eta(T_0)T_0^{-1}$ and $\rm{a_T^A}$ at low temperature is mainly originated from the swelling of $G(t)$ at shourt time scale ($t/{\rm{a_T^A}}<10\tau$), which is the consequence of the breakdown of TTS of $G(t)$. Therefore, assuming identical temperature dependence between $1/D$ and ${{\rm{a}_T}^A}$, the increasing of $D{\eta}T^{-1}$ as decreasing temperature as shown in the inset figure of Figure 7 can be explained from the viewpoint of the decreasing of ${\rm{a_T^A}}/{\rm{a_T^B}}$ at low temperature (Figure 3).  
\clearpage
\begin{figure}[h] 
\centering
  \includegraphics[height=10cm]{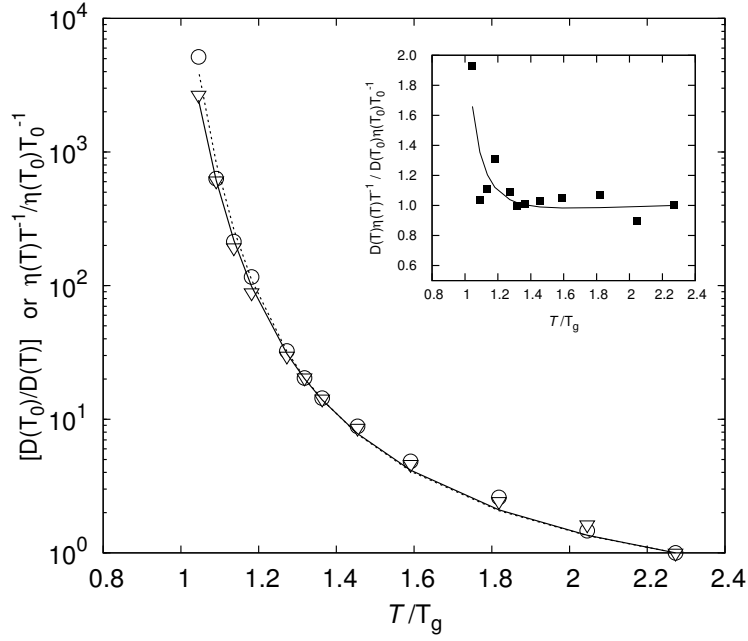}
  \caption{The $T_g$ scaled temperature dependence of $D(T_0)/D(T)$ and $\eta(T)T^{-1}/\eta(T_0)T_0^{-1}$ for $FJC$ melt. The open circles represent the values of $D(T_0)/D(T)$ and the open inverted triangles represent the value of $\eta(T)T^{-1}/\eta(T_0)T_0^{-1}$. Solid and dashed lines are least squares curves fit to the WLF equation (Eq.9) for $D(T_0)/D(T)$ and $\eta(T)T^{-1}/\eta(T_0)T_0^{-1}$, respectively. The inset figure shows the the $T_g$-scaled temperature dependence of the ratio ($D(T){\eta}(T)T^{-1}/D(T_0){\eta}(T_0)T_0^{-1}$). The filled squares represent the values calculated by means of MD simulations, and solid line represent the same quantity but using WLF-fitted values for $D(T_0)/D(T)$ and $\eta(T)T^{-1}/\eta(T_0)T_0^{-1}$.}
  \label{fgr:example}
\end{figure}
\clearpage

\subsection{E) Dynamic heterogeneity }
In order to explain the breakdown of the TTS of $G(t,T)$, the dynamic heterogeneity in the $FJC$ melt was investigated. In the equilibrium state, the motion of each particle in a liquid is a simple continuous stochastic process due to thermal fluctuations independent from other particle movements.
However, when the system temperature is near the $T_g$, instead of the continuous motion of an individual particle being frozen out, dynamic clusters of beads, the so-called cooperatively rearranged region ($CRR$), appeared. The existence of $CRR$ leads to the dynamics of an amorphous polymer melt becoming very heterogeneous. Based on some viscoelastic and dielectric measurements, such a heterogeneity would bring about a breakdown of the TTS of the relaxation process behavior in a polymer melt. The degrees of the dynamic heterogeneity can be evaluated using the self-part of the van Hove function, which is expressed by the followed equation:
\begin{equation}
G_{s}({\rm{r}},t)=\langle{\sum_{j}^{N}{\delta}\left({\rm{r}}(t)-\Delta{\rm{r}}_{j}(t)\right)}\rangle, 
\end{equation}
 where ${\rm{r}}$ is a position vector, $N$ is the number of beads, $j$ is an index of a bead, $\delta()$ denotes the "${\delta}$" function, and ${\rm{r}}_j$ and $\Delta{\rm{r}}_{j}$ are the position and displacement vectors of $j$-th bead, respectively. The $4{\pi}r^{2}G_{s}(t/{\rm{a}}_{T}{\sim}1{\tau},{\rm{d}}r)$ and $4{\pi}r^{2}G_{s}(t/{\rm{a}}_{T}{\sim}100{\tau},{\rm{d}}r)$ at $T$ = 1 and 0.48 as a function of ${\rm{d}}r$ are shown in Figure 7(upper) and 7(lower), respectively. The dashed lines in Fig.7 are the Gaussian distributions represented by Eq.10,
\begin{equation}
4{\pi}{r}^{2}G_s(r,t)=4{\pi}\left(\left({\frac{2}{3}}\right){\pi}d^2\right)^{-3/2}{r^2}{\rm{Exp}}\left({-\frac{3r^2}{2{d^2}}}\right),
\end{equation}
 where $d$ is the root-mean-square displacement of the beads, $d$ = 0.47${\sigma}$ ($t/{\rm{a}}_{T}{\sim}1{\tau}$) and $d$=2.1${\sigma}$ (${t}/{\rm{a}}_{T}{\sim}100{\tau}$). Both $G_{s}(t/{\rm{a}}_{T}{\sim}1{\tau}$) and $G_{s}(t/{\rm{a}}_{T}{\sim}100{\tau})$ at $T$ = 1 are in good agreement with each corresponding Gaussian distribution. This means that both bead motions for the short($\tau_A=1\tau$) and long($\tau_B=100\tau$) times at high temperature($T$ = 1) can be expressed as a simple stochastic process. However, when the system temperature ($T$ = 0.48) approaches its glass transition temperature ($T_g$ = 0.44), the bead motions on the short time scale($\tau_B = 1 \tau$) deviate from the Gaussian distribution, and exhibit a broader distribution. $G_{s}(t/{\rm{a}}_{T},{\rm{d}}r)$ at $T$ = 0.48 again becomes a Gaussian distribution on the long time scale($t/{\rm{a}}_{T} {\sim} 100 {\tau}$). In order to specify the duration of the heterogeneous period, the non-Gaussian parameter($\alpha_{2}\left(t\right)$) is evaluated from Eq.11.
\begin{equation}
\alpha_{2}\left(t\right)\equiv\frac{3\langle\Delta{r}^{4}\left({t}\right)\rangle}{5\langle\Delta{r}^{2}\left({t}\right)\rangle^{2}}-1.
\end{equation}
 The value of $\alpha_{2}\left(t\right)$ is a statistical value showing the degree of deviation from a Gaussian distribution. $\alpha_{2}\left(t\right)$ = 0 means that its distribution is equivalent to a Gaussian. The higher $\alpha_{2}\left(t\right)$, the more its distribution deviates from a Gaussian shape. 
Figure 8 shows $\alpha_{2}\left(t\right)$ at $T$ = 1 and $T$ = 0.48 as a function of the logarithmic time. $\alpha_{2}\left(t\right)$ at $T$ = 1 (filled triangles in Fig.8), which is sufficiently higher than the temperature in its glass transition region, is nearly zero on the plotted time scale ($t$ = 0.001 to 1000$\tau$). In contrast, $\alpha_{2}\left(t\right)$ at $T$ = 0.48 (open circles in Fig.8), which is near its $T_g$($T_g$ = 0.44), exhibits a maximum value around $t/{{\rm{a}}_T^{A}}$ = 0.4 $\tau$ and $\alpha_{2}\left(t\right)$ rapidly decreased to zero at times longer than $t/{{\rm{a}}_T^{A}}$ > 1$\tau$, which corresponds to the period for the $G(t, T=0.48)$ that begins to collapse to the master curve. This strongly suggested that the breakdown of TTS of $G(t,T)$ is related to its dynamic heterogeneity. From the experimental observations, it also suggests the relationships between the breakdown of TTS and its dynamic heterogeneity.\cite{Sokolov2009,Harrowell2012,Zorn1997}
\begin{figure}[h] 
\centering
  \includegraphics[height=10cm]{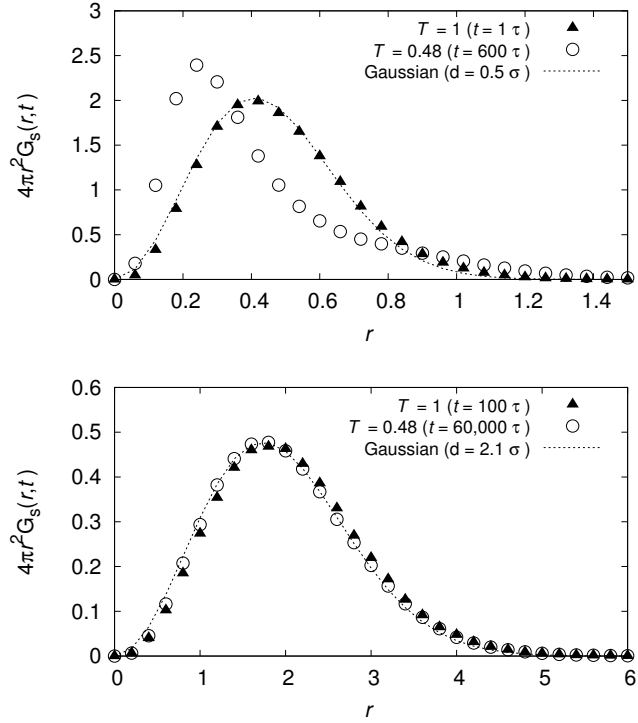}
  \caption{The self-part of the van Hove correlation function of $FJC$ melt at $T$ = 1 and $T$ = 0.48. The upper graph represents $G_{s}(r,dt)$ with dt $\approx$ $\tau_B$ $\approx$ $t/{\rm{a}}_T^A(T)$, for which the root mean square displacements of beads in melts at $T$ = 1 and at $T$ = 0.48 are approximately 0.5 $\sigma$. The lower graph represents $G_{s}(r,dt)$ with dt $\approx$ $\tau_A$ $\approx$ $t/{\rm{a}}_T^A(T)$, for which one of the beads in the melts at $T$ = 1 and at $T$ = 0.48 are approximately 2.1 $\sigma$. 
The broken lines in each graph are the Gaussian form (Eq.10) with $d$ = 0.47 $\sigma$ and 2.1 $\sigma$.}
  \label{fgr:example}
\end{figure}
\begin{figure}[h] 
\centering
  \includegraphics[height=5cm]{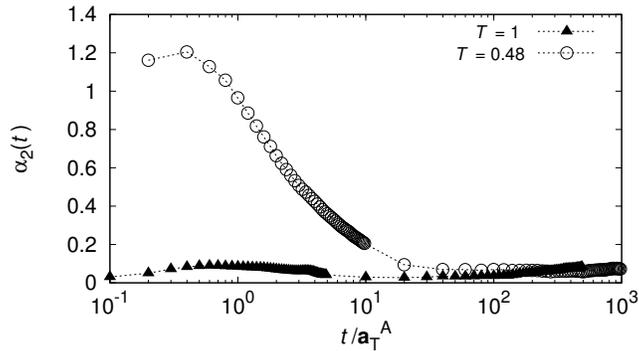}
  \caption{The non-gaussian parameters $\alpha_2$ of $FJC$ melt as a function of ${\rm{a}}_T^A(T)$-scaled time. The open circles represent the values of $\alpha_2$ at $T$ = 0.48, and the filled triangles represent the values of $\alpha_2$ at $T$ = 1.0. Dot lines are drawn point-to-point with a straight line with no smoothing.}
  \label{fgr:example}
\end{figure}
\clearpage

\section{Conclusion}
It has been in this study the breakdown of the time-temperature superposition (TTS) near its glass transition temperature ($T_g$) in simple bead-spring polymer melts with and without a chain angle potential. The stress relaxation modulus at different temperatures $G(t,T)$ are calculated by the Green-Kubo relations. The TTS of $G(t,T)$ in the bead-spring polymer melts works well at temperatures sufficiently higher than its $T_g$. However, when the system temperature is approaching the glass transition regime, a breakdown of the TTS is observed.
\\
 At temperatures near the $T_g$, the temperature dependence of the shift factor (${\rm{a}}_T^B$), which is defined on the time scale between the bond relaxation and the chain relaxation regimes of a $G(t)$-function, is significantly stronger than ones (${\rm{a}}_T^A$) defined on the time scale of the chain relaxation modes.The decoupling of the Stokes-Einstein law of diffusion-viscosity relation also appears in association with the breakdown of TTS of $G(t,T)$ in the glass transition regime. 
\\
The analysis of the van Hove function $G_{s}(r,t)$ and non-gaussian parameter $\alpha_{2}\left(t\right)$ of the bead motions strongly suggest that the breakdown of TTS is concerned with the dynamic heterogeneity. 
\\
The effect of the chain stiffness on the temperature dependence of the shift factors was also investigated in this study. The stiffer chain melt has a stronger temperature dependence of the shift factors than the ones of the flexible chain melt. However, regardless of the chain stiffness, the stress relaxation modulus functions of the bead-spring polymer melts will begin to breakdown the TTS at a similar temperature around $T \approx 1.2 T_g $. It is very interesting that such a simple bead-spring model (linear topology, without side-chain group) investigated in this study, exhibits the temperature-dependencies of the shift factors on the short and long time scales, which are qualitatively comparable to the broadband dielectric relaxation spectroscopy in several well-defined homo-polymers. 
 This unexpectedly implies that the simple polymer model, such as a bead-spring model, can be applicable to the investigation of the universal behaviors that appear in a supercooled polymer melt.

\begin{acknowledgement}
The author thanks Professor J.Takimoto at Yamagata University, Department of Polymer Science and Engineering, for the very valuable suggestions and advice. The author also thanks the members of the working group for computer chemistry of the Japan Association for Chemical Innovation, that deepens the utilization and application research of Integrated Simulator for Soft Materials; OCTA, for their useful discussions. 
\end{acknowledgement}


\begin{thebibliography}{50}
 
\bibitem{Daly2009}
J. Dealy, D. Plazek, \emph{Rheol.Bull.} \textbf{78} (2009) 16.

\bibitem{Ferry1980}
JD. Ferry, \emph{Viscoelastic Properties of Polymers} (1980), Wiley, New York. 

\bibitem{Plazek1965}
D. Plazek, \emph{J. Phys. Chem.} \textbf{69},3480 (1965).

\bibitem{Stillinger2001}
PG. Debenedetti and FH. Stillinger \emph{Natue} \textbf{410}, 259 (2001).

\bibitem{Inoue1999}
T.Inoue, T. Onogi,ML. Yao, K. Osaki, \emph{J. Polym. Sci. B Polym. Phys.} \textbf{37}, 389 (1999).

\bibitem{Roland1996}
PG. Santangelo, KL. Ngai, and CM. Roland, \emph{Macromolecules} \textbf{29}, 3651 (1996). 

\bibitem{Sokolov2006}
Y. Ding and AP. Sokolov, \emph{Macromolecules} \textbf{39}, 3322 (2006).

\bibitem{Sokolov2007}
AP. Sokolov and Y. Hayashi, \emph{Journal of Non-Crystalline Solids} \textbf{353}, 3838 (2007).

\bibitem{Plazek2001}
CM. Roland,KL. Ngai, PG. Santangelo, XH. Qiu, MD. Ediger, and DJ. Plazek, \emph{Macromolecules} \textbf{34}, 6159 (2001). 

\bibitem{Sokolov2009}
AP. Sokolov and KS. Schweizer, \emph{Physical Review Letters} \textbf{102}, 248301 (2009).

\bibitem{Harrowell2012}
MD. Ediger and P. Harrowell,\emph{The Journal of Chemical Physics} \textbf{137}, 080901 (2012).

\bibitem{Roland2013}
KL. Ngai and CM. Roland, \emph{The Journal of Chemical Physics} \textbf{139}, 036101 (2013).

\bibitem{Aoyagi2002}
T. Aoyagi, F. Sawa, T. Shoji, H. Fukunaga,J. Takimoto, and M. Doi, \emph{Computer Physics Communications} \textbf{145}, 267 (2002).

\bibitem{Nose_Hoover1}
S. Nos\'{e}, \emph{Molecular Physics: An International Journal at the Interface
  Between Chemistry and Physics}, \textbf{52}, 255 (1984). 

\bibitem{Nose_Hoover2}
W. G. Hoover, \emph{Phys. Rev. A}, \textbf{31}, 1695 (1985). 

\bibitem{Nose_Hoover3}
S. Nos\'{e}, \emph{The Journal of Chemical Physics}, \textbf{81}, 511 (1984). 

\bibitem{Andersen1980}
HC. Andersen, \emph{J. Chem. Phys.}, \textbf{72}, 2384 (1980). 

\bibitem{Buchholz2002}
J. Buchholz, W. Paul, F. Varnik, and K. Binder, \emph{J. Chem. Phys.}, \textbf{117}, 7364 (2002). 

\bibitem{Likhtman2007}
AE. Likhtman, SK. Sukumaran, and J. Ramirez, \emph{Macromolecules}, \textbf{40}, 6748 (2007).

\bibitem{KG1990}
K. Kremer and GS. Grest, \emph{The Journal of Chemical Physics}, \textbf{92}, 5057 (1990).

\bibitem{Roland1998}
PG. Santangelo and CM. Roland, \emph{Macromolecules}, \textbf{31}, 4581 (1998). 

\bibitem{Roland2003}
C.M. Roland and R. Casalini, \emph{J. Chem. Phys.}, \textbf{119}, 1838 (2003).

\bibitem{Schwartz2006}
G.A. Schwartz, J. Colmenero and A. Alegria, \emph{Macromolecules}, \textbf{39}, 3931 (2006).

\bibitem{Kunal2008}
K. Kunal, CG. Robertson, S. Pawlus, SF. Hahn, and AP. Sokolov, \emph{Macromolecules}, \textbf{41}, 7232 (2008). 

\bibitem{Swallen2003}
SF. Swallen,PA. Bonvallet, RJ. McMahon, and MD. Ediger, \emph{Phys. Rev. Lett.}, \textbf{90}, 015901 (2003). 

\bibitem{Urakawa2004}
O. Urakawa, SF. Swallen, MD. Ediger, ED. von Meerwall, \emph{Macromolecules}, \textbf{37}, 1558 (2004). 

\bibitem{Roland2004}
CM. Roland, KL. Ngai, and DJ. Plazek, \emph{Macromolecules}, \textbf{37}, 7051 (2004). 

\bibitem{Yamamoto1998}
R. Yamamoto and A. Onuki, \emph{Phys. Rev. Lett.}, \textbf{81}, 4915 (1998). 

\bibitem{Varnik2002}
F. Varnik and K. Binder, \emph{J. Chem. Phys.}, \textbf{117}, 6336 (2002). 

\bibitem{Zorn1997}
R. Zorn, \emph{Physical Review B}, \textbf{55}, 6249 (1997). 


\end{thebibliography}
\end{document}